\documentclass[sigconf,nonacm]{acmart}
\usepackage{tikz}
\usetikzlibrary{positioning,arrows.meta,fit,calc,shapes.geometric}
\usepackage{booktabs}
\usepackage{array}
\usepackage{xcolor}

\newcommand{\sys}{SignalCount}

\newcommand{\clf}{Jev}
\newcommand{\clfvendor}{Typesafe AI}

\begin{document}

\title{SignalCount: Approximate Counts with Structured and Semantic Predicates in Seconds}

\author{Matteo Agius-D'Arrigo}
\email{matteo@outersignal.com}
\affiliation{%
  \institution{OuterSignal}
  \city{New York}
  \state{New York}
  \country{USA}}

\author{Zachary Zelner}
\email{zach@outersignal.com}
\affiliation{%
  \institution{OuterSignal}
  \city{New York}
  \state{New York}
  \country{USA}}

\author{Harrison Ju}
\email{harrison@outersignal.com}
\affiliation{%
  \institution{OuterSignal}
  \city{New York}
  \state{New York}
  \country{USA}}

\author{Rushab Munot}
\email{rushab@outersignal.com}
\affiliation{%
  \institution{OuterSignal}
  \city{New York}
  \state{New York}
  \country{USA}}

\renewcommand{\shortauthors}{Agius-D'Arrigo et al.}

\begin{abstract}
Many applications store records with structured columns and a text document, and ask how many records in a filtered set meet a criterion that only a reader of the text can judge, such as ``How many patients with type 2 diabetes, aged 40 to 70, stopped metformin because their blood sugar stayed high?''
An interactive application needs this count in seconds, as the user types the criterion, to decide whether the matching set is worth materializing with a batch pipeline in which an LLM judges the records.
Prior approximate query processing over such semantic predicates has an expensive model, the oracle, label a sample at query time, so every query waits for the oracle.
We present \sys{}, a production system that answers COUNT queries with structured and semantic predicates and calls no oracle at query time, even for ad-hoc criteria.
Its oracle is the production LLM judge, and the judge's logged verdicts calibrate, offline, the fast classifier that already runs in front of the judge.
At query time, \sys{} counts the filtered set exactly and draws a uniform sample of at most 1,000 records in one columnar scan, scores the sample with the classifier, and multiplies the average calibrated probability by the exact count.
Its 90\% interval combines a sampling term with one-sided, rate-dependent margins for calibration error, fitted by split conformal prediction across tenants.
In a multi-tenant production deployment, latency has a p99 of 3.8\,s and stays flat as the filtered set grows beyond the sample.
On tenants held out from fitting, the interval covers the count that the production pipeline returns in 92\% of estimates, although estimates run about 15\% high in aggregate, because a calibration nearly unbiased per record is still biased per criterion.
\end{abstract}



\maketitle

\section{Introduction}
\label{sec:intro}

A research hospital is recruiting for a diabetes trial and needs patients whose first-line drug, metformin, failed to control their blood sugar.
A coordinator filters the patient records to type 2 diabetes, ages 40 to 70, seen this year and living near a trial site, and 50,000 patients match.
The filter is exact, because every part of it is a column.
The reason a patient stopped metformin is not a column: it is written in the clinical notes, keyword search cannot separate stopping for side effects from stopping because it stopped working, and no fixed taxonomy anticipates every trial's criterion.
So the coordinator writes the criterion in plain language, and a pipeline of models reads every patient's notes to find the matches.
That pipeline is the right tool for the final list, but it runs over the whole filtered set and takes long enough that nobody wants to run it to find out whether a criterion was worded well.

\begin{figure}[t]
  \centering
  \includegraphics[width=\columnwidth]{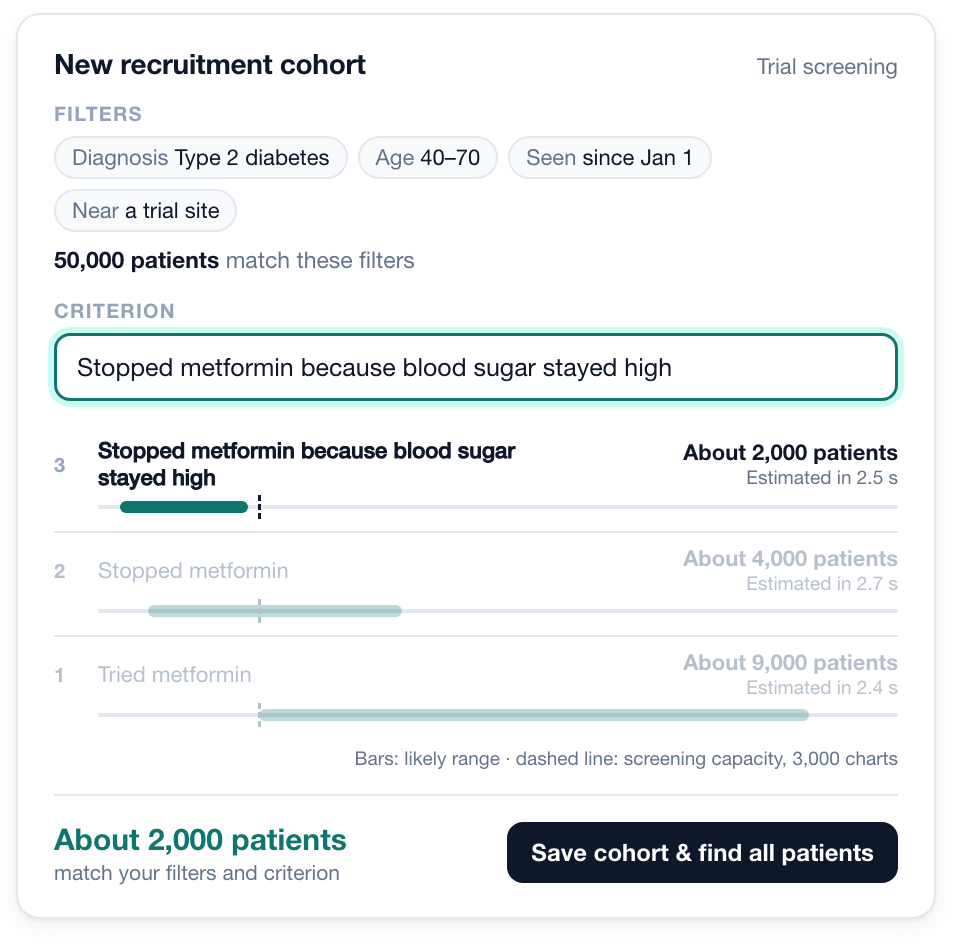}
  \Description{A cohort builder for trial recruitment. Filter chips read Diagnosis Type 2 diabetes, Age 40 to 70, Seen since Jan 1 and Near a trial site, and 50,000 patients match. The criterion box reads Stopped metformin because blood sugar stayed high. Below it, the versions of the criterion are listed newest first, each with its estimate and the time the estimate took: the current one, Stopped metformin because blood sugar stayed high, about 2,000 patients, estimated in 2.5 seconds; and, grayed out, Stopped metformin, about 4,000 patients in 2.7 seconds, and Tried metformin, about 9,000 patients in 2.4 seconds. Under each attempt, a bar shows its likely range against a dashed line at a screening capacity of 3,000 charts; only the current range lies below it. At the bottom, About 2,000 patients match your filters and criterion, next to a button Save cohort and find all patients.}
  \caption{Shaping a cohort as the user types the criterion. Each estimate arrives in seconds; bars show its likely range against the screening capacity (dashed). Saving runs the batch pipeline.}
  \label{fig:ui}
\end{figure}

And a criterion is rarely worded well the first time (Figure~\ref{fig:ui}).
``Tried metformin'' matches about 9,000 of those patients, and the coordinators can screen only 3,000 charts this quarter.
``Stopped metformin'' still sweeps in patients who stopped for stomach side effects.
``Stopped metformin because blood sugar stayed high'' is the cohort the trial wants.
The coordinator needs to see, while typing and within seconds, roughly how many patients the criterion selects and how sure that number is, so the cohort is shaped before the slow pipeline runs once.
The same loop appears wherever a costly action follows a count over records with text.
An insurer asks how many open property claims ``describe flood damage, not wind damage'' before sending inspectors, and a marketer how many support chats ``complain about a problem our new product solves'' before paying for a campaign.

\sys{} answers that question: an approximate COUNT over an exactly filtered set with one semantic predicate, returned with a 90\% interval within seconds.
For the third criterion it estimates about 2,000 matching patients with a 90\% interval of 418 to 2,808, in about 2.5\,s.
Both ends inform the decision: the high end fits the screening capacity, and the low end still leaves enough candidates to fill the trial, so the coordinator saves the cohort and the pipeline finds every match.
The example's numbers are illustrative; Section~\ref{sec:eval} evaluates the production deployment.

In the multi-tenant SaaS application where \sys{} runs, saving a criterion runs a batch pipeline that materializes the exact set with embedding and classifier filters, then a production LLM judge on the most promising records.
For a filtered set of over 500{,}000 records, judging every record this way would cost about 1{,}100 times as much as one estimate and take at least half an hour (Section~\ref{sec:eval-cost}).

\paragraph{Why existing methods wait.}
Prior work on aggregates with expensive predicates pairs a cheap model, the \emph{proxy}, with an expensive model, the \emph{oracle}, whose answer defines the truth.
ABae~\cite{kang2021abae} scores every record with the proxy, groups the records into strata by score, and has the oracle label a sample from each stratum at query time.
The oracle's labels give the estimate and its interval, and the proxy only decides where to sample; later systems follow the same pattern~\cite{russo2023inquest,dai2024uqe,hui2026olla}.
Prediction-powered inference (PPI)~\cite{angelopoulos2023ppi} instead averages the proxy's predictions over the records and corrects that average by the proxy's error on a small sample that the oracle labels.
Both intervals hold however poor the proxy is, and both pay for that at query time: each new predicate needs fresh oracle labels, and the proxy runs on every record.
In our deployment the oracle is a production LLM judge, and a single call to it has a p50 of 3.7\,s and a p90 of 7.7\,s.
A sample labeled fully in parallel finishes no sooner than its slowest call, which is beyond a budget of 5\,s, and the proxy pass grows with the filtered set.
We argue this from the judge's measured latency and did not run either method in production.

\paragraph{No oracle labels at query time.}
\sys{} moves the oracle offline.
The production pipeline already runs a fast non-generative classifier in front of its judge (\clf{}, Section~\ref{sec:arch}), which scores how likely a record's text is to meet a natural-language criterion, and logs the classifier's score and the judge's verdict for every record the judge sees.
The classifier is the proxy, and those logs are oracle labels that cost nothing more to collect.
Once calibrated to those verdicts, the classifier's probabilities on a small uniform sample can answer the query directly.
The calibration must then transfer to the queried criterion and filtered set, so the interval must account for calibration error.
AlloyDB moves proxy training offline for known queries~\cite{chung2026proxy}, and calibrated classifiers measure prevalence with no LLM call for fixed categories in a daily batch~\cite{xu2026calibrate}; \sys{} does so for arbitrary criteria posed at query time, over an exactly counted filtered set, in seconds.

\paragraph{Requirements.}
\textbf{(R1) Latency.} A p99 below 5\,s for filtered sets from a handful of records to the largest a tenant has, with model calls and model cost per query that do not grow with the filtered set.
\textbf{(R2) A validated interval.} Every estimate carries a nominal 90\% interval, also when the criterion is rare and the sample holds no match, and its empirical coverage is measured overall and by estimated rate.
\textbf{(R3) Agreement with the production judge.} The target is the number of records the production pipeline marks as matching once the criterion is saved.

\paragraph{Approach.}
The database gives the denominator exactly, and only the fraction of matches is estimated.
A single analytical query returns the exact size of the filtered set and a seeded uniform sample of at most 1{,}000 record ids.
The classifier scores their documents, an isotonic table fitted on logged (score, verdict) pairs turns each score into the probability that the judge would accept the record, and the estimate is the average probability times the exact size.
Its 90\% interval widens the estimate by two one-sided margins for calibration error, fitted by split conformal prediction and varying smoothly with the estimated rate, and by a finite-population sampling term.
Each statistical component is established; the contribution is their composition into a path that calls no oracle at query time and whose model work is bounded by the sample cap, an interval whose calibration margins are fitted across tenants, and the measurement of its coverage on held-out tenants in production.

\par
\begin{figure}[t]
  \centering
  \includegraphics[width=\columnwidth]{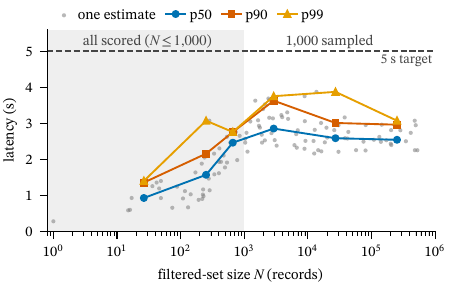}
  \Description{Scatter plot of end-to-end latency against filtered-set size on a log axis for 112 estimates, with percentile lines per size group and a dashed 5 second line. Latency rises with size up to 1,000 records, where the whole set is scored, and stays flat near 2.5 to 3.9 seconds above that, below the 5 second line.}
  \caption{Latency of 112 production estimates against the size of the filtered set. Sets of at most 1,000 records are scored in full (shaded). Lines join percentiles per size group.}
  \label{fig:teaser}
\end{figure}

\ifdim\dimexpr\pagegoal-\pagetotal\relax<4\baselineskip\newpage\fi
\paragraph{Contributions.}
\begin{itemize}
  \item \textbf{A latency-bounded design} for COUNT queries with structured and semantic predicates, whose model work is bounded by the sample cap (Sections~\ref{sec:arch} and~\ref{sec:est}).
  \item \textbf{One-sided, rate-dependent calibration margins}, fitted by split conformal prediction across tenants, which with a refitted table raise held-out coverage from 85.5\% to 92.0\% over a single symmetric margin, at a narrower median width (Sections~\ref{sec:est-margins} and~\ref{sec:eval-heldout}).
  \item \textbf{An evaluation on the production deployment} (Section~\ref{sec:eval}). On repeated tenant splits of 2{,}119 estimates across hundreds of tenants, the nominal 90\% interval covered the true count in 92.0\% of held-out estimates. Over 112 estimates against tenants' saved filters, latency had a p99 of 3.76\,s (Figure~\ref{fig:teaser}), and the stages bounded by the sample cap stayed near-constant even for sets of millions of records.
\end{itemize}

\section{Problem}
\label{sec:problem}

\paragraph{Data and query.}
A \emph{tenant} is one customer organization of the application, and it has a table of \emph{records}: one person each in our deployment, one patient each in the running example.
A \emph{scope} selects the records a query may touch, such as the tenant's records after deduplication and exclusions.
The structured columns of each record live in a columnar store (ClickHouse), and the structured predicates form a filter compiled to SQL.
For records the offline pipeline has processed, a text document (JSON in our deployment) lives in a row or document store keyed by record id.
Only records with a document can be judged, so the \emph{filtered set} is the records of the scope that pass the filter and have a document, and $N$ is its exact size.
The query is a COUNT over the filtered set with one semantic predicate, a criterion in the form the production judge uses.
In the running example, the scope is the hospital's patient records and the filter and criterion are those of Figure~\ref{fig:ui}.

\paragraph{The judge and the target.}
For a criterion, the production pipeline gives each record a yes-or-no verdict; in the terms of prior work, its classifier is the proxy and its judge the oracle.
Its last two stages are a small classifier and an LLM judge.
The classifier scores each record between 0 and 1 for the criterion; only records scoring at least the routing threshold of 0.10 reach the judge, whose verdict is final, and the rest receive a no.
Earlier, cheaper stages decide which records the classifier sees.
The target, the \emph{true count}, is the number of records in the filtered set whose production verdict under the saved criterion is yes, with records the pipeline has not evaluated counted as no (R3).
The estimator observes scores, not verdicts, so we model each verdict given its score as a coin flip whose probability depends on the score, and the interval \sys{} returns is a \emph{prediction interval} for the true count, not a confidence interval for a parameter.

\paragraph{Problem statement.}
Given a tenant, a criterion, a filter and a scope, \sys{} returns $N$, an estimate of the true count and an interval at a \emph{nominal} level of 90\%, the level the interval claims.
Its \emph{empirical} coverage is the measured fraction of queries whose interval contains the true count.
That fraction should be about 90\%, counted over queries (tenant, filter and criterion), with the randomness in the sample and in the verdicts given the scores.
The p99 latency must be below 5\,s.
Rows fetched by key, model calls and model cost per query must be bounded by constants independent of $N$, so that the columnar scan of the scope is the only step whose work grows with the data.
We judge the system by the coverage of its interval and the error of its estimate against the true count; it outputs no per-record labels.

\paragraph{Scope.}
The target is the production judge, not human labels; how well the judge agrees with independent labels is outside this paper.
The calibration is specific to one classifier and one judge, and the table and margins must be refitted when either changes.
Documents are produced offline, so an estimate is as fresh as the pipeline and covers only records with a document.
\sys{} answers COUNT and rate over one semantic predicate, with one classifier family and one judge, in one production deployment.

\section{System Architecture}
\label{sec:arch}

\begin{figure*}[t]
  \centering
  \begin{tikzpicture}[
    box/.style={draw, rounded corners=2pt, align=center, font=\footnotesize, minimum height=1.15cm, text width=2.7cm, fill=white},
    store/.style={draw, cylinder, shape border rotate=90, aspect=0.18, align=center, font=\scriptsize, minimum height=0.75cm, text width=1.9cm, fill=gray!8},
    arr/.style={-{Stealth[length=2mm]}, thick}]
    \node[box] (route) at (0,0) {\textbf{API route}\\auth, role, rate limit,\\estimate cache};
    \node[box] (ch) at (3.7,0) {\textbf{Analytical query}\\counts, bottom-k sample\\{\scriptsize\itshape\color{blue!55!black} 812 / 1045 ms}};
    \node[store] (chdb) at (3.7,-1.5) {columnar store};
    \node[box] (eb) at (7.6,0) {\textbf{Document fetch}\\parallel keyed reads\\{\scriptsize\itshape\color{blue!55!black} 481 / 1365 ms}};
    \node[store] (rstore) at (7.6,-1.5) {row store};
    \node[box] (score) at (11.1,0) {\textbf{Scoring}\\one \clf{} call per record,\\bounded in flight\\{\scriptsize\itshape\color{blue!55!black} 1753 / 2294 ms}};
    \node[box] (est) at (14.6,0) {\textbf{Estimator}\\calibrate, estimate,\\90\% interval};
    \draw[arr] (route) -- (ch);
    \draw[arr] (ch) -- (chdb);
    \draw[arr] (ch) -- (eb);
    \draw[arr] (eb) -- (rstore);
    \draw[arr] (eb) -- (score);
    \draw[arr] (score) -- (est);
    \draw[arr] (est.south) -- ++(0,-1.85) -| (route.south)
      node[pos=0.25, above, font=\scriptsize] {estimate, interval, flags, counts};
  \end{tikzpicture}
  \Description{A left-to-right pipeline of five boxes: API route, analytical query on a columnar store, document fetch from a row store, scoring with one classifier call per record, and the estimator, which returns the estimate, interval, flags and counts to the route. The query, fetch and scoring boxes are labeled with p90 and p99 latencies of 812/1045, 481/1365 and 1753/2294 ms.}
  \caption{Request path of an estimate. Every stage after the analytical query does work bounded by the sample cap of 1,000. Labels give p90 / p99 per stage over the latency workload (Section~\ref{sec:eval-latency}).}
  \label{fig:arch}
\end{figure*}
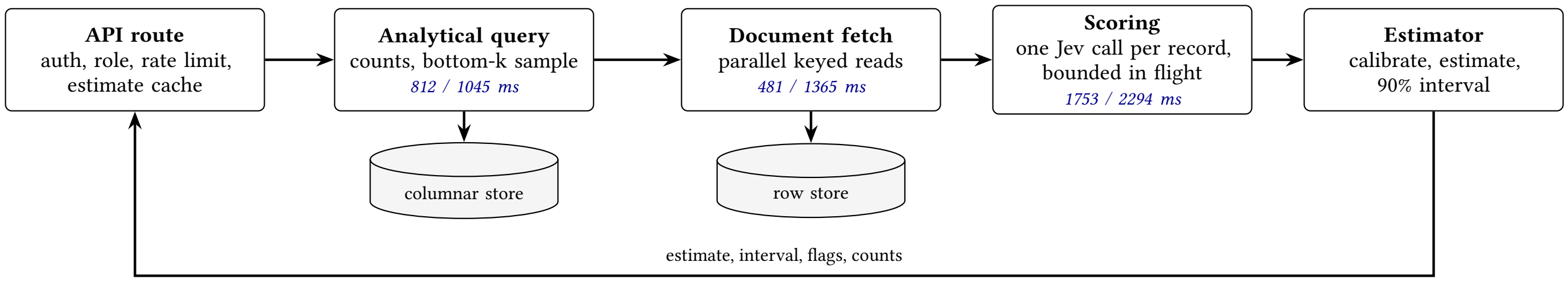

Figure~\ref{fig:arch} shows the request path, which crosses three services: the web application's API route, the document service that owns the document store, and the scoring service.
An estimate writes no production state, only cache entries, a rate-limit counter, and log and trace records.
Only the analytical query touches the whole scope.
Every later stage does work bounded by the sample cap, so its rows fetched, model calls and model cost do not grow with $N$.

\subsection{Counting and Sampling in One Scan}
\label{sec:arch-count}
The estimate needs the exact size of the filtered set and a uniform sample of it.
The count must match what the tenant sees in the filter preview.
A separate sampling query would read the filtered set a second time.
Taking the first rows the store returns would make the sample depend on how ids are laid out.

We therefore draw the sample inside the aggregate that computes the count.
The structured predicates are compiled to SQL by the compiler that the filter preview uses.
A single query returns the exact filtered count, the scope's count and the sample:
\begin{verbatim}
SELECT count(*) FILTER (WHERE <filter>)   AS N,
       count(*)                           AS N_U,
       TOP_K_BY(1000, hash(id, <seed>), id)
         FILTER (WHERE <filter>)          AS ids
FROM records WHERE <scope> AND <has document>
\end{verbatim}
Here \texttt{TOP\_K\_BY} stands for an ordered-array aggregate that keeps the 1,000 filtered records with the smallest seeded hash, a bottom-k sample~\cite{cohen2007bottomk}.
If the hash acts as a random permutation of ids, the sample is uniform without replacement.
When $N$ is at most 1,000, the sample is the whole set.
For the running example, the query returns $N = 50{,}000$ and 1,000 patient ids.

The seed is a hash of the compiled query and its parameters.
Repeated estimates therefore agree, an edit to the filter draws a fresh sample, and a retry cannot average away an unlucky draw.
Uniformity becomes a property of the hash rather than of the id layout.
The design removes a second read of the filtered set and adds only a bounded heap to the scan.
Any engine with an ordered top-k or array aggregate can do the same.

\subsection{Fetching Documents by Key}
The document service receives the criterion, the sampled ids and $N$.
Documents are large JSON values, and one serial read of the whole sample would dominate the path.
The service therefore splits the ids into chunks and reads them in parallel by key.
Each chunk is one read-only lookup by key, restricted to the requesting tenant, of the form \texttt{WHERE org\_id = \$1 AND id = ANY(\$2)}.

If any chunk fails, the whole read fails as retryable.
A partial sample would be biased toward the chunks that finished, so the service never returns one.
The fetched documents are shuffled, so that the request-size cap, which keeps a prefix, drops a random subset.

\subsection{Scoring as a Bounded Fan-Out}
\label{sec:arch-score}
The judge is too slow to call at query time.
The scoring service instead scores every sampled record with a fast non-generative classifier, a model that answers structured questions about an input without generating text.
It is \clf{} (\clfvendor{})\footnote{\url{https://docs.typesafe.ai/introduction}}, which the production pipeline uses to decide which records reach its judge.
An embedding cannot take the classifier's place: it encodes each document before the criterion is known, so it captures topic but misses the negations and fine distinctions that criteria turn on, and its similarity has no common scale across criteria to calibrate once.
For each record, the service asks whether the criterion is met for the record's document, built with the pipeline's prompt construction, one criterion per request.
\clf{} answers with a score between 0 and 1, which Section~\ref{sec:est} calibrates to the judge's probability of accepting the record.
Any non-generative decision model that returns a score can replace \clf{}, including self-hosted open-weight models, so scoring throughput is a matter of provisioning.

The classifier sits behind the production pipeline's gateway, and an estimate must not starve the pipeline.
The service issues one request per record through a process-wide limiter on calls in flight.
Each attempt has a 1.7\,s timeout, with at most two attempts and a short backoff between them.
Each service admits a bounded number of estimates and refuses the rest with a retry hint.
The scoring service aborts outstanding calls when its caller disconnects.
As a result, an admitted estimate keeps its bounded latency and a refused one fails fast.

The service maps each score through the calibration table.
It returns the estimate, its interval on the rate and on the count, and the number of records scored and dropped.
It sets a \emph{rare} flag when the estimated rate is below 2\%, and a \emph{below-threshold} flag when no sampled record reached the routing threshold.
The route adds $N$, the scope's count and the scope-wide rate.

\subsection{Caching and Cost}
Every uncached estimate makes one metered classifier call per sampled record, and a user editing a criterion repeats queries.
The route rate-limits estimates per tenant.
It caches each response for 600\,s, keyed by tenant, application context, criterion text and a hash of the compiled query.
Any edit to the filter therefore misses the cache.
The gateway serves no cached answers to estimate calls, so every uncached estimate pays for its own scores.
The cost of an estimate is thus a constant set by the sample cap; Section~\ref{sec:eval-cost} measures it against judging every record.

\subsection{Failure Handling}
Each hop classifies a failure by whether an identical retry could succeed.
\emph{Transient} errors are a taken slot, a store, read or service timeout, or too many sample losses (Section~\ref{sec:est-rare}).
They return 503 with \texttt{retryable: true}, or 429 with \texttt{Retry-After} when a slot is busy, and the interface retries them.
\emph{Permanent} errors mean a contract between services has changed or a credential is wrong.
They return 502, are logged at error level and are never retried.
Timeouts bound every request: the query timeout above plus a 20\,s document-service timeout.

The document service checks every scoring response against invariants that any correct estimate satisfies.
All fields must be present, and the low end, estimate and high end must be in order within $[0, N]$.
The rates must lie within $[0, 1]$, at least one record must be scored, and the drop count must be non-negative.
The route also requires the response to echo the $N$ it estimated over.
Each answered request logs one structured line with per-stage times, and the path emits no per-record logs.

\section{Estimation}
\label{sec:est}

As in ABae~\cite{kang2021abae}, correctness is separate from efficiency.
When $N$ exceeds the sample cap, coverage depends on calibration and on the uniformity of the sample, and latency depends on the cap.

\subsection{Calibration}
\label{sec:est-cal}
A monotone table maps each classifier score to the probability that the judge accepts the record.
We fit it by weighted isotonic regression~\cite{zadrozny2002calibration} over score bins of width 0.01, from the routing threshold of 0.10 to 1.00.
Below the threshold, where the judge is never consulted, the probability is zero.
The table has 91 entries and is a constant in the scoring service.
The \emph{deployed calibration}'s table is fitted, with no model calls, on logged pairs of a query-time score and the production verdict for the same record and criterion, from the main study of Section~\ref{sec:eval}.
It replaced an \emph{initial calibration} with a single symmetric margin.

\begin{table}[t]
  \caption{Expected calibration error of the calibration table. Out of fold: refitted on 5 tenant folds of the main study's pairs, each pair scored by the table fitted without its tenant. Bottom: reliability of the initial table on the main study's pairs.}
  \label{tab:ece}
  \footnotesize
  \setlength{\tabcolsep}{3pt}
  \begin{tabular}{@{}l r r r@{}}
    \toprule
    Table, pairs & Pairs & All & $\ge$ 0.10 \\
    \midrule
    Initial, in-sample study & 188{,}757 & 0.0065 & 0.0262 \\
    Initial, main study & 1{,}660{,}862 & 0.0083 & 0.0275 \\
    Deployed form, main study, out of fold & 1{,}660{,}862 & 0.0037 & 0.0012 \\
    \midrule
    Calibrated probability & Records & Predicted & Observed \\
    \midrule
    0 (score below 0.10) & 1{,}373{,}189 & 0 & 0.0042 \\
    $[0, 0.1)$ & 20{,}165 & 0.0736 & 0.0573 \\
    $[0.1, 0.2)$ & 50{,}064 & 0.1316 & 0.1043 \\
    $[0.2, 0.3)$ & 35{,}125 & 0.2381 & 0.2007 \\
    $[0.3, 0.4)$ & 26{,}510 & 0.3435 & 0.3131 \\
    $[0.4, 0.5)$ & 20{,}833 & 0.4453 & 0.4186 \\
    $[0.5, 0.6)$ & 16{,}938 & 0.5507 & 0.5265 \\
    $[0.6, 0.7)$ & 13{,}516 & 0.6593 & 0.6290 \\
    $[0.7, 0.8)$ & 15{,}771 & 0.7533 & 0.7241 \\
    $[0.8, 0.9)$ & 17{,}285 & 0.8553 & 0.8166 \\
    $[0.9, 1.0]$ & 71{,}466 & 0.9770 & 0.9548 \\
    \bottomrule
  \end{tabular}
\end{table}

We measure the table by its expected calibration error (ECE): the gap between the calibrated probability and the observed acceptance rate, averaged over bins and weighted by each bin's share of records (Table~\ref{tab:ece}).
We use one bin for scores below the routing threshold, whose probability is zero, and 10 equal-width bins on the calibrated probability.
The bin below the threshold holds 82.7\% of the main study's pairs and dominates the figure over all records, so we also report ECE over the records at or above the threshold.
Refitted out of fold by tenant on the main study's 1.66 million pairs, the deployed form has an ECE of 0.0037 over all records and 0.0012 at or above the threshold, against 0.0083 and 0.0275 for the initial table.
A small error pooled over records does not certify a count, because errors of opposite sign across criteria cancel (Section~\ref{sec:eval-bias}).

Coverage rests on what we call the \emph{calibration assumption}.
For every record of the filtered set, (a) the judge accepts it with the probability the table gives its query-time score, (b) verdicts are independent given the scores, and (c) they are independent of which records are sampled.
The design does not rely on the assumption holding exactly.
The table is fitted across tenants and criteria but applied to one criterion within one filtered set, and calibration on average does not imply calibration on a subset.
The margins therefore widen the interval for this failure, and they cover per-criterion error only as well as their fit transfers.
Section~\ref{sec:eval-heldout} measures the resulting coverage on held-out tenants.

\subsection{The Estimate}
Number the records of the filtered set $1$ to $N$.
Let $Y_i$ be 1 if the judge accepts record $i$ and 0 otherwise, and let $p_i$ be the calibrated probability the table gives record $i$'s score.
The true count $\theta$ is the sum of the verdicts.
The estimated rate $\bar p$ is the average of $p_i$ over the $n$ scored records of the sample, and the estimate $\hat\theta$ is that average times the exact size $N$:
\begin{equation}
  \label{eq:est}
  \theta = \sum_{i=1}^{N} Y_i, \qquad
  \bar p = \frac{1}{n} \sum_{i \in \text{sample}} p_i, \qquad
  \hat\theta = N \bar p .
\end{equation}
The service reports $\hat\theta$ rounded to a whole number.
This is probabilistic classify-and-count~\cite{bella2010quantification}, a variant of classify-and-count~\cite{forman2008quantifying}, on a uniform sample.

Under the calibration assumption the estimate has zero expected error given the scores, because $\hat\theta$ and $\theta$ have the same expected value, the sum of $p_i$ over the filtered set.
For $\hat\theta$, a uniform sample's mean is unbiased for the set's mean~\cite[Thm.~2.1]{cochran1977sampling}, so $\bar p$ averages to $\frac{1}{N}\sum_{i=1}^{N} p_i$.
For $\theta$, part (a) of the assumption says each $Y_i$ is 1 with probability $p_i$.
The estimate needs no record-level accuracy.
It does need calibration on the filtered set (Section~\ref{sec:eval-bias}) and accounting for every sampled record removed before scoring.
The seed is fixed per query, so coverage is a property over queries, not over repeats of one query.
In the running example, 50 of the 1,000 sampled patients have a calibrated probability of 0.8 and the rest score below the threshold.
The average is 0.04, and the estimate is 2,000 patients.

\subsection{The Interval}
\label{sec:est-interval}
The nominal 90\% interval should miss the true count about 5\% of the time on each side.
Two sources of error separate the estimate from the true count, and they call for different tools.
Sampling error comes from scoring $n$ records instead of all $N$; textbook survey statistics give its size, and it shrinks as $n$ grows.
Calibration error comes from applying a table fitted across tenants and criteria to one criterion; it is the same however many records are scored, and no formula gives its size, so we measure it on past criteria whose true count is known.
We size the first with a \emph{sampling term} $T$, derived below on the supposition that the table is right, and the second with a \emph{margin} for when it is not.
Table~\ref{tab:notation} lists the quantities used.

\begin{table}[t]
  \caption{Quantities in the interval, and when each is set.}
  \label{tab:notation}
  \footnotesize
  \begin{tabular}{@{}lp{3.5cm}p{3.1cm}@{}}
    \toprule
    Quantity & Meaning & Set \\
    \midrule
    $N$ & size of the filtered set & per query, counted exactly \\
    $n$ & records scored, at most 1{,}000 & per query \\
    $Y_i$, $\theta$ & verdict on record $i$ (1 or 0); true count, their sum & unknown at query time \\
    $p_i$ & record $i$'s calibrated probability & per record, read from the table \\
    $\bar p$, $\hat\theta$ & estimated rate; estimate $N\bar p$ & per query, from the sample \\
    $V$, $W$ & variance of $p_i$; average of $p_i(1-p_i)$ & per query, from the sample \\
    $T$ & sampling term & per query, from $N$, $n$, $V$, $W$ \\
    Table & score to probability, 91 entries & offline, shared by all tenants \\
    $f_L$, $f_U$ & low and high factors (Table~\ref{tab:margins}) & offline, across tenants; read at the estimated rate \\
    1.645, 2.303 & normal 95th percentile; $-\ln 0.10$ & constants \\
    \bottomrule
  \end{tabular}
\end{table}

\paragraph{If the table is right, two things are still left to chance.}
Suppose the calibration assumption holds.
The estimate still differs from the true count, for two unrelated reasons.

The first is that only a sample is scored.
Write $\mu$ for the average of $p_i$ over all $N$ records of the filtered set.
Scoring every record would give the estimate $N\mu$.
The service has $N\bar p$ instead, and $\bar p$ differs from $\mu$ by the luck of which records were drawn.

The second is that a probability is not a verdict.
A record with $p_i = 0.8$ is accepted four times in five, but any one such record is either accepted or not.
So even $N\mu$, the expected number of accepted records, differs from the actual number $\theta$, just as 100 tosses of a coin that lands heads four times in five rarely give exactly 80 heads.

The error of the estimate is the sum of these two gaps, because $N\mu$ cancels:
\begin{equation}
  \label{eq:split}
  \hat\theta - \theta =
  \underbrace{\big( N \bar p - N\mu \big)}_{\substack{\text{sample gap: which}\\\text{records were drawn}}}
  \; + \;
  \underbrace{\big( N\mu - \theta \big)}_{\substack{\text{verdict gap: how}\\\text{the verdicts fell}}} .
\end{equation}

\paragraph{The size of each gap.}
We measure the size of a chance error by its variance, the square of its standard deviation.
We first write each variance in terms of all $N$ records, as if every probability were known.
Two of the quantities involved, $V_{\text{all}}$ and $W_{\text{all}}$, are not known at query time, and the next paragraph replaces them with their values on the sample.

\emph{Sample gap.}
The average of $n$ values drawn at random from a population has a variance equal to the population's variance divided by $n$.
Write $V_{\text{all}}$ for the variance of $p_i$ over all $N$ records.
Because the sample is drawn without replacement from a finite set, the variance is also multiplied by $1 - n/N$~\cite[Thm.~2.2]{cochran1977sampling}: a sample that covers most of the set leaves little to chance, and one that covers all of it leaves nothing.
The gap is $N$ times the error of the average, and multiplying a quantity by $N$ multiplies its variance by $N^2$.
The variance of the sample gap is therefore $N^2 (1 - n/N)\, V_{\text{all}} / n$.

\emph{Verdict gap.}
This gap is $\sum_{i=1}^{N} (p_i - Y_i)$, one term per record.
A verdict that is 1 with probability $p_i$ has variance $p_i(1-p_i)$, which is zero for a record that is certain either way and largest, at 0.25, for a record at 0.5.
By part (b) of the assumption the verdicts are independent, and the variances of independent quantities add.
Write $W_{\text{all}}$ for the average of $p_i(1-p_i)$ over all $N$ records, so that the sum of the $N$ variances is $N\, W_{\text{all}}$.
This gap does not involve the sample: it remains when every record is scored, because the verdicts are still unknown.

By part (c), which records are drawn says nothing about how the verdicts fall.
The two gaps are therefore independent, and their variances add:
\begin{equation}
  \label{eq:var}
  \begin{aligned}
    \mathrm{Var}(\hat\theta - \theta)
    &= N^2 \Big(1 - \frac{n}{N}\Big) \frac{V_{\text{all}}}{n} + N\, W_{\text{all}} \\
    &= N^2 \bigg[ \Big(1 - \frac{n}{N}\Big) \frac{V_{\text{all}}}{n} + \frac{W_{\text{all}}}{N} \bigg] .
  \end{aligned}
\end{equation}
The second line only factors $N^2$ out of the first.

\paragraph{From the variance to the sampling term.}
Equation~\ref{eq:var} cannot be computed as written: $V_{\text{all}}$ and $W_{\text{all}}$ are taken over all $N$ records, and only the sample is scored.
The same two quantities computed on the sample, $V$ (with divisor $n-1$) and $W$, are on average equal to them~\cite[Thm.~2.4]{cochran1977sampling}, so they take their place.
Substituting them into Equation~\ref{eq:var} and taking the square root gives the standard deviation of the error.
The error is a sum of many small contributions, so it is approximately normally distributed~\cite{hajek1960clt}, and a normal quantity falls more than 1.645 standard deviations above its mean 5\% of the time, and as often that far below.
The sampling term is therefore 1.645 standard deviations:
\begin{equation}
  \label{eq:T}
  T = 1.645\, N \sqrt{\Big(1 - \frac{n}{N}\Big) \frac{V}{n} + \frac{W}{N}} .
\end{equation}
If the table is right, the true count lies within $T$ of the estimate about 90\% of the time.

In the running example the sample holds 50 probabilities of 0.8 and 950 of zero, so $V = 0.0304$ and $W = 0.008$.
The sample gap has variance $50{,}000^2 \times 0.98 \times 0.0304 / 1{,}000$, about 74{,}500, a standard deviation of 273 patients.
The verdict gap has variance $50{,}000 \times 0.008 = 400$, a standard deviation of 20 patients.
Together the variance is about 74{,}900 and the standard deviation 274 patients, so $T = 1.645 \times 274$, about 450 patients.
The sample gap is 99.5\% of the variance.

\paragraph{If the table is wrong: the margin.}
The sampling term supposes the table is right.
For one criterion it may not be: the table may give that criterion's records probabilities that are, say, a third too high.
The estimate is then a third too high as well, whatever the sample size.
This error is a factor, not an amount, and scoring more records does not shrink it.
The margin therefore multiplies the estimate, by a low factor below 1 in case the table runs high for this criterion, and by a high factor above 1 in case it runs low.
No formula gives the factors.
Section~\ref{sec:est-margins} fits them on past criteria with known true counts, and they depend on the estimated rate.

\begin{figure}[t]
  \centering
  \includegraphics[width=\columnwidth]{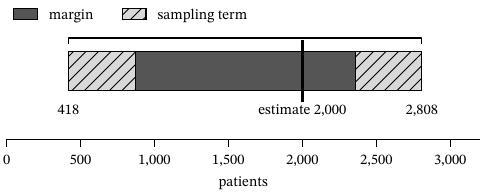}
  \Description{A horizontal bar on a patient-count axis showing the interval from 418 to 2,808 around an estimate of 2,000. Each side is split into a calibration-margin segment and a sampling segment of about 450 patients, and the low-side margin segment is the widest.}
  \caption{The running example's 90\% interval around the estimate of 2,000 patients: the calibration margin (wider on the low side) and the sampling term on each side. Of the width of 2{,}390, the margins contribute about 1{,}490 and the sampling term about 900; the floor of about 115 patients does not bind.}
  \label{fig:range}
\end{figure}

\paragraph{The interval.}
The interval applies the margin first and then the sampling term on each side (Figure~\ref{fig:range}):
\begin{align*}
  \text{low} &= \text{estimate} \times \text{low factor} - T, \\
  \text{high} &= \max\big(\text{estimate} \times \text{high factor} + T,\; 2.303\,N/n\big) .
\end{align*}
Here the estimate is $\hat\theta$ before rounding.
The low end is rounded down and clipped at zero, and the high end rounded up and clipped at $N$, so the interval always contains the estimate.
The floor $2.303\,N/n$ keeps the high end of a sample with only a few weak matches at or above the bound that a sample of $n$ records with no match supports (Section~\ref{sec:est-rare}).

In the running example, at an estimated rate of 4\%, the factors of Section~\ref{sec:est-margins}, interpolated between the centers of Table~\ref{tab:margins}, are a low factor of 0.43 and a high factor of 1.18.
The low end is $2{,}000 \times 0.43 - 450$ and the high end $2{,}000 \times 1.18 + 450$; with unrounded factors the interval is 418 to 2,808 patients.

\subsection{Fitting the Margins}
\label{sec:est-margins}

The initial calibration widened every estimate by one symmetric margin.
On held-out tenants it covered fewer estimates than the nominal level claims.
Its misses were one-sided: almost all were overestimates, with the truth below the low end, and they fell at the upper estimated rates (Section~\ref{sec:eval-fitting}).

The deployed margins are therefore one-sided and depend on the estimated rate.
Write $f_L$ for the low factor and $f_U$ for the high factor at the estimated rate.
In symbols, before rounding and clipping, the interval of Section~\ref{sec:est-interval} is
\begin{equation}
  \label{eq:interval}
  \begin{aligned}
    \text{low} &= \hat\theta\, f_L - T, \\
    \text{high} &= \max\big( \hat\theta\, f_U + T,\; 2.303\,N/n \big) .
  \end{aligned}
\end{equation}

\paragraph{What one past run needs.}
A calibration run is a past estimate whose true count $\theta$ is known.
For each run we ask how much widening would have been just enough for the interval to reach $\theta$.
On the low side, the low end $\hat\theta f_L - T$ must lie at or below $\theta$.
Setting it equal to $\theta$ and solving for the factor gives the run's \emph{needed low factor}.
The high side is the same with the signs reversed:
\begin{equation}
  \label{eq:needed}
  f_L^{\text{need}} = \min\Big( 1,\; \frac{\theta + T}{\hat\theta} \Big),
  \qquad
  f_U^{\text{need}} = \max\Big( 1,\; \frac{\theta - T}{\hat\theta} \Big) .
\end{equation}
Both use the run's own sampling term $T$.
A factor of 1 means no widening.
A run needs a low factor below 1 only when its estimate exceeds the true count by more than the sampling term, and a high factor above 1 only when it falls short by more.
The needed high factor is also 1 when the floor already reaches $\theta$.
For example, a run with an estimate of 2{,}000, a sampling term of 450 and a true count of 1{,}000 needs a low factor of $1{,}450 / 2{,}000 = 0.725$: scaled to 1{,}450, the estimate is within 450 of the truth.
In practice we search a fine grid of factors (steps of about 0.3\%) with the rounding applied, which agrees with Equation~\ref{eq:needed} up to the grid step.

\paragraph{From many runs to one factor.}
We group runs into four \emph{estimated-rate groups} (Table~\ref{tab:margins}).
Within each group of $m$ runs, we order the runs from the least widening needed to the most and take the factor needed by the run at position $\lceil 0.95\,(m+1) \rceil$.
About 95\% of the group's runs need that much widening or less, so about 5\% would be missed on that side.
This is the split-conformal rule~\cite{lei2018conformal} for a 5\% miss rate on each side, and so 90\% two-sided when runs are exchangeable.
Runs from one tenant are not exchangeable, and interpolation leaves the per-group rule, so coverage is measured rather than guaranteed.
The rule assumes no particular distribution for the calibration error.

Each factor sits at its group's center, the midpoint of the group's edges on a log scale (0.3\% for the lowest group, which has no lower edge), and between centers the service interpolates, so the bounds do not jump at group edges.
It interpolates the logarithm of each factor linearly in $\log_{10}$ of the estimated rate, and holds the factors constant outside the outer centers.
Both scales are logarithmic because both quantities are compared by ratio: 1\% to 10\% is as far as 10\% to 100\%, and a factor of 2 is as far from 1 as a factor of $1/2$.
The running example's rate of 4\% lies 39\% of the way from the 2.24\% center to the 10\% center on that scale, so its low factor lies 39\% of the way from 0.47 to 0.38, at 0.43.
We check that the bounds are non-decreasing in the estimated rate.
The lowest group's own high factor is close to 1, so it is raised to the high factor fitted on all runs together.
The factors are fitted on out-of-fold estimates of the main study's runs, with folds by tenant and each run estimated with a table fitted without its fold.
The two sides differ because estimates run high at low and middle rates.

\begin{table}[t]
  \caption{Deployed margin factors by estimated-rate group.}
  \label{tab:margins}
  \footnotesize
  \begin{tabular}{@{}lrrrr@{}}
    \toprule
    Estimated rate & below 1\% & 1--5\% & 5--20\% & above 20\% \\
    \midrule
    Group center & 0.3\% & 2.24\% & 10\% & 44.7\% \\
    Low factor $f_L$ & 0.81 & 0.47 & 0.38 & 0.53 \\
    High factor $f_U$ & 1.31 & 1.11 & 1.30 & 1.53 \\
    \bottomrule
  \end{tabular}
\end{table}

With a table refitted on query-time scores, these margins raise held-out coverage from 85.5\% to 92.0\% at a narrower median width (Section~\ref{sec:eval-heldout}).

\subsection{Rare Criteria, Small Sets and Losses}
\label{sec:est-rare}

\paragraph{Zero-event bound.}
For a rare criterion, the sample often holds no record at or above the threshold.
The estimate and the sampling term are then both zero, and an interval of zero to zero would claim a certainty the sample cannot support.
An empty sample does not show that no record matches.
It shows that matches are too rare to turn up in $n$ draws, and that puts a ceiling on the rate.

Suppose a fraction $r$ of the filtered set matches.
One record drawn at random is then not a match with probability $1-r$.
Two records drawn independently are both not matches with probability $(1-r)(1-r)$, and $n$ records with probability $(1-r)^n$.
The higher the rate, the less likely an empty sample: at a rate of 1\%, a sample of 1{,}000 comes up empty with probability $0.99^{1000}$, about 1 in 23{,}000, so after an empty sample a rate of 1\% is not credible.
We take as the upper bound the rate at which an empty sample still has a 10\% chance, since any higher rate makes what we observed less likely than that:
\begin{equation}
  \label{eq:zero}
  (1 - r)^n = 0.10
  \;\;\Longrightarrow\;\;
  r = 1 - 0.10^{1/n} \le \frac{2.303}{n} .
\end{equation}
The last step replaces the exact solution by a simpler value that is never smaller: 2.303 is the natural logarithm of 10, and for $n = 1{,}000$ both are 0.23\%.
In the running example's set of 50{,}000 patients that is 115 patients, the floor noted in Figure~\ref{fig:range}.
Drawing without replacement makes an empty sample slightly less likely than $(1-r)^n$, so the bound errs on the safe side.
This is the ``rule of three'' at the 90\% level~\cite{hanley1983zero}.
The service then returns an estimate and low end of zero, and a high end at the rate $\max(0.01, 2.303/n)$, capped at 1.
It sets the rare and below-threshold flags, so that an interface can show ``fewer than'' the high end.
For $n$ above 230 the 0.01 floor dominates, so the bound holds at a level above 90\%, increasingly so toward the sample cap; for smaller $n$ it holds at a one-sided 90\%, not the 95\% of the other intervals.
The bound covers the true count only if every record with a positive verdict also scores at or above the threshold at query time.
That holds for nearly all records (Section~\ref{sec:eval-rare}).

\paragraph{Small filtered sets.}
When $N$ is at most the sample cap, the whole filtered set is scored (a census), and the first part of the sampling term vanishes.
The interval then reflects verdict variability, the margins, and the floor when the rate is small.
The document service sets $N$ to the number of listed records it read, so a record that left the filtered set between the two reads is not counted.

\paragraph{Minimum sample and losses.}
The service refuses an estimate that scored fewer than 100 records, or fewer than $N$ when $N$ is below 100.
Records can be lost before scoring: rows no longer complete, and the request-size cap.
The document service counts these losses where the sample is drawn.
A rejected classifier request drops only its record.
Two limits apply, each at 20\% of the sample.
If documents over the classifier's size limit exceed it, the estimate is refused as unusable, a permanent error.
If total losses, including transient classifier errors, exceed it, the estimate is refused as degraded, a transient error.
Dropped records are treated as missing at random.

\section{Evaluation on the Production Deployment}
\label{sec:eval}

We evaluate latency (R1), then the interval's coverage and the estimate's bias on held-out tenants (R2, R3), baselines, rare criteria and cost.

\subsection{Deployment and Workloads}
\label{sec:eval-setup}

\paragraph{Deployment.}
\sys{} is generally available in a production application that serves thousands of tenants, where it backs custom filters and criteria for audience segmentation.
The route runs in serverless functions and the two services on containers.
The authors issued every estimate in this section through that deployment, against real tenants' data.

\paragraph{Workloads.}
The \emph{latency workload} is 112 estimates on non-empty saved filters of a small set of tenants, each with its tenant's saved criterion, issued with a 1.5\,s gap.
The \emph{in-sample study} is 244 estimates over 142 filtered sets from fewer than 25 tenants, all in the fit of the initial calibration's margin.
Of its estimates, 125 use a tenant's own saved filter, 51 the whole scope with no filter, and 68 a filter saved by another tenant.
The \emph{main study} carries our main accuracy result, on tenants outside every earlier fit.
We compiled every active saved filter of every active tenant and computed the true count for every eligible criterion: active, evaluated by the production pipeline, and created at least 30 days before the run.
From each stratum of four true match-rate groups by four size groups, we drew up to 150 (filtered set, criterion) pairs, with at most 6 per tenant per stratum and 3 criteria per filter.
We excluded every tenant of the in-sample study and added 60 pairs with a true count of zero.
Of the 2{,}123 estimates issued across hundreds of tenants, 2{,}119 were answered: 2{,}059 with a positive true count and 60 with zero.
Another 300 estimates, 294 answered, repeated 30 of these filtered sets under 10 sample seeds; the estimate's median coefficient of variation across seeds was 0.083.
The filtered sets reach 857{,}662 records in the main study and 1.7 million in the in-sample study.
Table~\ref{tab:data} summarizes the three workloads.

\begin{table}[t]
  \caption{The three workloads. Size and predicate rows count each issued filtered set once. True-count rows count answered estimates and leave out the in-sample study's 23 check pairs.}
  \label{tab:data}
  \footnotesize
  \setlength{\tabcolsep}{3pt}
  \begin{tabular}{@{}l r r r@{}}
    \toprule
     & Latency & In-sample & Main \\
    \midrule
    Tenants & fewer than 25 & fewer than 25 & hundreds \\
    Estimates issued / answered & 112 / 112 & 244 / 244 & 2{,}123 / 2{,}119 \\
    Filtered sets & 112 & 142 & 1{,}687 \\
    Criteria & 14 & 136 & 1{,}766 \\
    Estimates per tenant, median / max & 7 / 20 & 13 / 19 & 1 / 19 \\
    $N$, median & 2{,}435 & 5{,}561 & 2{,}362 \\
    $N$, p90 & 144{,}538 & 324{,}327 & 111{,}031 \\
    $N$, max & 527{,}197 & 1{,}697{,}655 & 857{,}662 \\
    True rate if positive, median / p90 & 4.8 / 77.8\% & 4.1 / 37.1\% & 2.8 / 30.9\% \\
    Estimates with a true count of 0 & 4 & 12 & 60 \\
    Structured predicates, 0 / 1 / 2+ & 12 / 55 / 33\% & 23 / 43 / 34\% & 38 / 49 / 14\% \\
    (Score, verdict) pairs & -- & 188{,}757 & 1{,}660{,}862 \\
    \quad with a positive verdict & -- & 11.7\% & 8.9\% \\
    \bottomrule
  \end{tabular}
\end{table}

\paragraph{What coverage means here.}
The deployed table and margins are fitted on the main study, so Section~\ref{sec:eval-heldout} re-runs the whole fitting procedure on 200 tenant splits and tests each on tenants outside both fits.
We report empirical coverage overall, by estimated rate and by side.
It is an average over held-out tenants and criteria, not a per-query guarantee.
It is also approximate: runs are exchangeable at best across tenants, and the study, like the margins' fit, stratifies on the outcome rather than following the mix of production queries.

\paragraph{True count and reproducibility.}
The true count (Section~\ref{sec:problem}) comes from the production verdicts without new model calls, and the input criterion is the saved criterion itself.
On 23 check pairs, the production-label rate has a median ratio of 0.975 to the judge's acceptance rate in a separate record of the same judge's verdicts.
The classifier gateway records every classifier call.
A line-for-line port of the deployed estimator, run on those scores, reproduces the estimate and both bounds exactly for 112 of 112 latency runs, 243 of 244 in-sample runs and 2{,}119 of 2{,}119 held-out runs.
We therefore evaluate every calibration variant below offline on the recorded scores.
The data cannot be released, because it consists of tenants' records, criteria and documents.

\paragraph{Failures.}
Of the 2{,}779 estimates issued in all studies, 2{,}769 were answered, and no transient error or timeout occurred.
In each of the other 10, the classifier rejected one record's request, which then failed the whole estimate; the deployed service drops and counts such a record (Section~\ref{sec:est-rare}).

\subsection{Latency}
\label{sec:eval-latency}

\paragraph{Tail latency.}
We measure latency from criterion to estimate: the analytical query, issued through a read-only proxy, plus the document-service call, which times its fetch and scoring stages.
It stays below the 5\,s target: over the latency workload, criterion-to-estimate latency had a p99 of 3.76\,s (Figure~\ref{fig:teaser}).
Scoring is the largest stage at every percentile, at a p50 of 1483\,ms of a 2491\,ms total over all 112 runs; Figure~\ref{fig:arch} gives p90 and p99 per stage.
Over the 71 filtered sets above 1,000 records, scoring took a median of 1.6\,s.
The stages bounded by the sample cap stay flat.
Over the 2{,}413 document-service calls of the main study and its repeats, that call had a p50 of 2.1\,s and a p90 of 2.6\,s.
Its p99 was 4.5\,s and its maximum 7.2\,s; scoring alone had a p99 of 3.5\,s, so the longer tail comes from the classifier.

\paragraph{Size of the filtered set.}
Latency does not grow with the filtered set once it exceeds the sample (Figure~\ref{fig:teaser}).
Below 1,000 records the set is scored in full, and latency grows with it.
For 1k--10k, 10k--100k and above 100k records, the p99 is 3.76, 3.88 and 3.08\,s, up to the workload's largest filtered set of 527{,}197 records.
Over the 71 sets above 1,000, the rank correlation of $N$ with latency is \textminus0.27.
On larger sets too, over the 179 in-sample estimates above 1,000 records, the document-service call had a p90 of 2.6 to 3.0\,s in every size group, up to 1.7 million records.

\paragraph{Size of the tenant's data.}
The analytical scan is the one stage that grows, with the scope rather than with $N$, and over the measured scopes it stays under a second.
Over 132 saved filters of the largest tenants, with scopes up to 1.7 million records, the analytical query had a p99 of 860\,ms (minimum of 3 runs, including the proxy hop).
On the largest scope, the scope boundary alone (deduplication and exclusions, no filter) took 251\,ms, and a filter on a parsed array field took 766\,ms, so evaluating that filter was about two thirds of the scan.
The scan is an ordinary filtered COUNT with a bounded top-$k$, so it scales as the columnar store does, while every other stage stays bounded by the sample cap.

\subsection{In-Sample Results with the Initial Calibration}
\label{sec:eval-fitting}

The in-sample study shows how a calibration misses, and why held-out tenants are needed.
On it, the initial calibration covered 191 of 209 positive-truth estimates, or 91.4\% (tenant-clustered bootstrap [86.7\%, 95.7\%]).
But every one of those tenants, and 82 of its 121 (tenant, criterion) pairs, were in the fit of its margin.
A table fitted on the in-sample tenants and fixed before any held-out estimate was issued covered only 87.0\% of the held-out positive-truth estimates.

Misses were one-sided.
Of the 18 misses, 17 were overestimates, with the truth below the low end: a lower miss rate of 8.1\% against an upper one of 0.5\%.
They fell in the two upper estimated-rate groups, whose coverage was 85.5\% and 81.8\%.
The margin provides most of the coverage.
Without it (both factors equal to 1), coverage fell to 54.3\%, and to 70.9\% on the 55 filtered sets scored in full, which have no sampling error.
A split-conformal margin fitted on other tenant folds gave 89.1\% [84.3\%, 92.6\%], with per-fold coverage from 75.4\% to 95.5\%.

\subsection{Held-Out Accuracy}
\label{sec:eval-heldout}

\begin{figure}[t]
  \centering
  \includegraphics[width=\columnwidth]{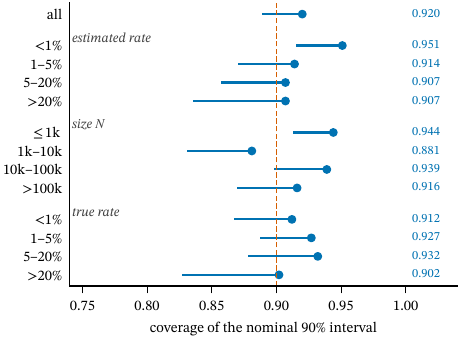}
  \Description{Dot plot of the deployed calibration's interval coverage, overall and by estimated rate, filtered-set size and true rate, with dots at the mean over 200 tenant splits and lines to the 5th percentile, against the nominal 90 percent. The deployed calibration's mean is 0.920 overall and between 0.90 and 0.951 in every group except sets of 1k to 10k records at 0.881.}
  \caption{Coverage on held-out tenants over 200 repeated tenant splits, by group, for the deployed calibration: mean (dot and number) and 5th percentile (line end).}
  \label{fig:heldout}
\end{figure}

\paragraph{Coverage on held-out tenants.}
On held-out tenants, the deployed calibration meets the nominal level on average.
We repeat the whole fitting procedure on 200 random 40/30/30 splits of the main study's tenants.
The table is fitted on the first part and the margins on the second, and the third is tested, so no test tenant contributes to either fit (Figure~\ref{fig:heldout}).
Coverage is 92.0\% on average, with a standard deviation of 1.7 points and a 5th percentile of 89.0\%.
Every group mean is at or above 90.2\%, except filtered sets of 1k--10k records, at 88.1\%.
The lowest 5th percentiles are 83.2\% for that group and 82.8\% for true rates above 20\%.

\paragraph{From the initial to the deployed calibration.}
The one-sided, rate-dependent margins account for most of the gain over the initial calibration.
The initial calibration used a table of the same form, fitted on 340{,}683 of the pipeline's own (score, verdict) pairs logged over four weeks.
Its one symmetric margin, a high factor of 1.43 and a low factor of 0.70 at every rate, was 1.645 robust standard deviations (0.216 each) of the per-criterion natural-log ratio of actual to predicted acceptances.
That ratio was taken over 645 (tenant, criterion) sets, whose fit contained every tenant of the in-sample study.
On the same 200 splits, the initial calibration covered 85.5\% of the held-out estimates.
The deployed calibration refits the table on query-time scores and uses the margins of Table~\ref{tab:margins}.
It raised coverage to 92.0\%, with every estimated-rate, size and true-rate group at or above the initial calibration's coverage.
Its median width is also narrower: 1.59 times the true count, against 1.64.
Refitting the table alone, with the initial margin, reached only 86.5\% under the cross-fitting below.

\paragraph{Choice of margin form.}
One-sided margins fitted per group covered 90.4\% to 90.5\% in every estimated-rate group, at a median width of 1.50 against 1.79 for a symmetric margin (Table~\ref{tab:alternatives}).
We compared forms by 5-fold cross-fitting over tenants, with the table and every margin fitted on other folds.
A single global pair of one-sided margins covered only 80.7\% in the top estimated-rate group.
Per-group quantiles fitted on overlapping calibration sets make per-group coverage close to an identity, however, so the repeated split above is our held-out measurement.
The deployed form adds log-rate interpolation, the floor on the lowest group's upper margin, and one table fitted on all tenants with margins from out-of-fold estimates (Section~\ref{sec:est-margins}).

\begin{table}[t]
  \caption{Design alternatives, evaluated out of fold. Top: intervals on the main study, cross-fitted by tenant. Bottom: point estimates on the 208 positive-truth in-sample estimates with reproduced scores, 5 tenant folds, initial calibration; ratio of sums by true rate.}
  \label{tab:alternatives}
  \footnotesize
  \setlength{\tabcolsep}{3pt}
  \begin{tabular}{@{}lrrrrr@{}}
    \toprule
    Interval & \multicolumn{2}{r}{Coverage} & \multicolumn{2}{r}{Worst group} & Width \\
    \midrule
    Global symmetric margin & \multicolumn{2}{r}{90.0\%} & \multicolumn{2}{r}{--} & 1.79 \\
    Global one-sided pair & \multicolumn{2}{r}{90.3\%} & \multicolumn{2}{r}{80.7\%} & -- \\
    One-sided per group & \multicolumn{2}{r}{90.5\%} & \multicolumn{2}{r}{90.4\%} & 1.50 \\
    EM prior shift + margins & \multicolumn{2}{r}{$\le$86.3\%} & \multicolumn{2}{r}{--} & 7.63 \\
    \midrule
    Point estimate & Med.\ err. & $<$1\% & 1--5\% & 5--20\% & $>$20\% \\
    \midrule
    Averaging (\sys{}) & 29.6\% & 1.45 & 1.58 & 1.12 & 0.92 \\
    ACC & 52.5\% & 0 & 0.50 & 1.06 & 1.05 \\
    PACC & 83.8\% & 0 & 0.63 & 1.00 & 1.07 \\
    Count $\ge$ 0.35 (fitted) & 21.6\% & -- & -- & -- & -- \\
    \bottomrule
  \end{tabular}
\end{table}

\subsection{Remaining Bias}
\label{sec:eval-bias}

\paragraph{Bias of the point estimate.}
The point estimate is biased high in aggregate, most at estimated rates of 1--5\%.
Over the repeated splits, the sum of estimates is 1.15 times the sum of true counts (5th to 95th percentile 1.07 to 1.27).
By estimated-rate group, the median signed relative error is +19.1\%, +29.5\%, +11.6\% and \textminus6.0\%.
The lower margins are fitted to this error (Section~\ref{sec:est-margins}), so coverage holds.
Section~\ref{sec:eval-ablation} compares the error with bias-correcting baselines.

\paragraph{Per-record and per-criterion calibration.}
The bias comes from how calibration error falls across criteria, not from the record-level fit.
Averaged over records, a calibration fitted on other data is nearly unbiased.
Over the 1.66 million (query-time score, production verdict) pairs of the main study, the initial calibration's probabilities sum to 1.01 times the positive verdicts.
Above the threshold every probability bin over-predicts by 0.02 to 0.04, while the records scored below it, counted as zero, hold 3.95\% of the positive verdicts.
With the initial calibration applied once to the full main study, the ratio of sums is 2.26 for criteria matching under 1\% of the filtered set and 0.90 for those over 20\%.
Binned by estimated rate, it is 1.42 at 1--5\%, 1.34 at 5--20\% and 0.93 above 20\%, so it is not regression to the mean.
Out of fold under the deployed calibration, runs whose true rate is below the pooled rate of their table's fit are overestimated (median signed error +28.7\%, ratio of sums 1.38).
Runs above it are underestimated (\textminus7.1\%, 0.91).
The log of a run's rate over the pooled rate and the log of its estimate over the truth have a Spearman correlation of \textminus0.46.
This is the shrinkage toward the pooled rate that prior work predicts~\cite{card2018calibration,keith2018uncertainty}, though it alone does not separate prior shift from other causes.
Ratios of sums weigh the largest filtered sets most.

\paragraph{Prior-shift correction.}
The standard prior-shift correction over-corrects in both directions, so we do not deploy it.
It re-estimates each query's prior by expectation maximization (EM) and rescales the calibrated probabilities to it~\cite{saerens2002adjusting}.
Cross-fitted by tenant on the main study, it brings the ratio of sums to 0.96 but more than doubles the median absolute error (66.7\% against 28.7\%).
In the lowest estimated-rate group the ratio of sums falls from 1.09 to 0.05 (median signed error \textminus100\%), and in the top group it rises from 0.89 to 1.15.
With conformal margins, its coverage is at most 86.3\%, at a median width of 7.63 times the truth (Table~\ref{tab:alternatives}).

\paragraph{Unseen criteria.}
The deployed calibration also covers criteria that no fit had seen.
All evaluated criteria were saved at least 30 days earlier, so they existed during the fit windows.
Of the held-out positive-truth estimates, though, 249 (124 tenants) have a tenant and criterion text that no earlier fit had seen.
With its fitting procedure re-run 5-fold by tenant, in an analysis fixed in advance, the deployed calibration covered 92.0\% [87.9\%, 95.5\%] of them out of fold.
Its lower and upper miss rates were 1.6\% and 6.4\%.

\subsection{Baselines and Ablations}
\label{sec:eval-ablation}

\paragraph{Quantification baselines.}
Averaging calibrated probabilities has the lowest median error of the three unthresholded estimators (Table~\ref{tab:alternatives}).
Adjusted classify-and-count (ACC)~\cite{forman2008quantifying} corrects a thresholded count with the threshold's true- and false-positive rates.
Its probabilistic variant (PACC)~\cite{bella2010quantification} corrects the mean calibrated probability likewise.
We fit both on 5 tenant folds of the in-sample study (ACC at a threshold of 0.35) and compare their point estimates out of sample with averaging under the initial calibration.
Both clip to zero for every estimate with a true rate below 1\% and undercount at 1--5\%.
Above a true rate of 5\%, they are within 8\% of the truth in aggregate.
Below 5\%, averaging over-predicts but has the lower median absolute error: 52.7\% and 34.1\%, against 63.7\% to 100\%.

\paragraph{Sample size.}
The median error is set by calibration rather than by sampling.
Offline on the recorded scores of 149 in-sample runs above 1,000 records, subsampling to $n = 500$ scored records left the median absolute error at 29.5\%, against 29.3\% at $n = 1{,}000$.
The p90 of scoring time falls from 1,788 to 1,286\,ms, so the cap of 1,000 still fits the latency target.
The larger sample is bought for the interval: the sampling term shrinks as $1/\sqrt{n}$ on large sets, and the zero-event bound $2.303/n$ halves (Section~\ref{sec:est-rare}).

\paragraph{Thresholding.}
Counting scores at or above a threshold instead of averaging gives a lower median absolute error at the threshold fitted on other folds (21.6\% at 0.35), but the error depends on the threshold (157.5\% at 0.1, 88.3\% at 0.9) and the interval under-covers: 0.889 against 0.913 for averaging, and 77.5\% above a true rate of 20\%.
So \sys{} averages calibrated probabilities.

\subsection{Rare Criteria}
\label{sec:eval-rare}

The interval holds for rare criteria.
For the 60 held-out pairs whose true count is zero, the initial calibration's interval contained 0 in 57 (95.0\%).
Under the deployed calibration, the lowest estimated-rate group covers 95.1\% on average (Figure~\ref{fig:heldout}).
On the in-sample study, the zero-event bound fired in 7 estimates with true counts of 2 to 71 records and covered all of them.
The bound assumes that matching records score at or above the threshold, and nearly all do: in the main study, 99.58\% of the 1.37 million sampled records scored below the threshold carry a negative production verdict.

The choice of the lowest group's high factor (Section~\ref{sec:est-margins}) does not drive this result: re-choosing it inside each of the 200 tenant splits leaves that group's held-out coverage at 95.2\%, against 95.1\% for the deployed choice.

\subsection{Cost against Judging Every Record}
\label{sec:eval-cost}

An estimate costs a constant set by the sample cap, while exhaustive scoring grows linearly with the filtered set.
The classifier gateway recorded 140 estimate executions of 1,000 records each over three days before the studies.
An estimate used about 2.0 million prompt tokens (upper median, range 1.7 to 2.5 million), or about 2,000 per call.
The exhaustive alternative runs the classifier on every record of the filtered set and the judge on every record at or above the threshold, 20.0\% of records in the in-sample reliability data.
We price both at list model prices, with the classifier's measured tokens per record on both sides and the judge's mean tokens per call over two days of production calls.
For 1,000 records, exhaustive scoring costs about 2.1 times as much as an estimate; the difference is the judge.
It costs 21 times as much for 10,000 records, 210 times for 100,000, and 1,100 times for a filtered set of over 500,000 records.
For the largest set, the judge alone would take 30.9 minutes at its peak observed production throughput, and 83.8 hours at its median.

\section{Lessons Learned}
\label{sec:lessons}

\paragraph{Holding out by tenant.}
Coverage measured on the tenants a calibration was fitted on overstated what new tenants get, by about six percentage points for our initial calibration (Sections~\ref{sec:eval-heldout} and~\ref{sec:eval-fitting}).
For a calibration shared across tenants, the unit to hold out is the tenant, and the test needs many tenants.

\paragraph{Per-criterion bias.}
A calibration nearly unbiased when pooled over records can still be biased per criterion (Section~\ref{sec:eval-bias}): a count that averages calibrated probabilities inherits the prior of the criteria the table was fitted on~\cite{keith2018uncertainty,xu2026calibrate}, so reliability diagrams pooled over records cannot certify it, and the check must be made per criterion, on the filtered sets being estimated.

\paragraph{One-sided, rate-dependent margins.}
Our calibration error was neither symmetric nor constant across rates (Section~\ref{sec:eval-heldout}), and margins fitted per estimated-rate group addressed both.
Coverage comes mostly from the margin, as binomial-only intervals for classify-and-count are known to undercover~\cite{keith2018uncertainty}, while doubling the sample from 500 to 1,000 barely changed the error (Sections~\ref{sec:eval-fitting} and~\ref{sec:eval-ablation}).

\paragraph{Prior-shift correction.}
The standard answer to this bias, EM re-estimation of each query's prior~\cite{saerens2002adjusting}, fixed our aggregate while more than doubling the median absolute error (Section~\ref{sec:eval-bias}).
EM assumes the same within-class score distribution for every criterion, which need not hold when the judge reads each record's text~\cite{card2018calibration}; the over-correction in both directions is consistent with that mechanism.
A bias correction should be checked per estimated-rate group before it ships, because the aggregate it fixes can hide errors in both directions.

\paragraph{Semi-structured predicates.}
Once the semantic predicate is bounded by the sample cap, the cost that grows is the scan, and within it the structured predicates that read fields inside JSON text (Section~\ref{sec:eval-latency}).
Fields that filters read are candidates for typed columns materialized from the semi-structured source.

\paragraph{Applicability.}
The approach needs an exact count for the structured predicates, a fast classifier for the semantic predicate, and logged pairs of the classifier's score and the expensive model's verdict.
Model cascades, in which a cheap model screens rows in front of an expensive LLM, are common practice~\cite{liskowski2026aisql,chung2026proxy,liskowski2026cascades}.
Such a cascade already computes a proxy score and an oracle verdict for every row it routes to the oracle, and logging the pair is all calibration needs.

\section{Related Work}
\label{sec:related}

\paragraph{Approximate query processing.}
Sampling-based AQP answers aggregates with error bounds, from online aggregation~\cite{hellerstein1997online} to stored samples~\cite{agarwal2013blinkdb,chaudhuri2007stratified} and query-time samplers in production~\cite{kandula2019aqp}.
Chaudhuri et al.\ observe that AQP research has yet to reach products~\cite{chaudhuri2017silver}.
In AQP the scan is the expensive part; in \sys{} the semantic predicate is, so the scan evaluates the structured predicates exactly and the sample serves only the semantic one.

\paragraph{Proxy-based query processing.}
Kang et al.\ use a cheap proxy model to reduce calls to an expensive oracle, for aggregation, selection with guarantees and semantic indexes~\cite{kang2019blazeit,kang2020supg,kang2022tasti}.
ABae~\cite{kang2021abae} and InQuest~\cite{russo2023inquest} run the proxy on every record, stratify on its score, and have the oracle label a sample at query time.
Their intervals need no assumption about the proxy, but every query waits for the oracle; \sys{} trades those labels for the calibration assumption (Section~\ref{sec:est-cal}).
BlazeIt's query rewriting is one of the few systems that, like ours, calls no oracle at query time: it labels data at ingestion and bounds error from those labels.
But it trains a specialized network per object class and assumes no drift, while our criteria are arbitrary text posed at query time and our margins are fitted across tenants and measured on held-out ones.
Chung et al.\ replace LLM filters in BigQuery and AlloyDB with proxies trained on LLM labels, offline in AlloyDB for known queries~\cite{chung2026proxy}.
The proxy filters every row and gives no interval on an aggregate; \sys{} serves arbitrary criteria and scores only a capped sample.

\paragraph{Semantic operators and LLM-powered engines.}
LOTUS~\cite{patel2025lotus} and BARGAIN~\cite{zeighami2026bargain} guarantee output quality relative to a gold model, and Cortex AISQL~\cite{liskowski2026aisql} runs proxy-oracle cascades in production with precision-recall guarantees on a filter's output.
For aggregates, UQE~\cite{dai2024uqe} estimates COUNT without an interval from LLM labels on an embedding-stratified sample, and ThalamusDB~\cite{jo2024thalamusdb} and OLLA~\cite{hui2026olla} bound aggregates over multimodal data and text; all three call an expensive model at query time.
SemStats~\cite{xu2026semstats} is closest in problem: it estimates semantic-filter cardinality~\cite{zimmerer2026semceb,urban2026semhist} over unstructured data.
An LLM builds an index per corpus and maps each query onto it, and a judge labels a sample.
Unlike all of these, \sys{} adds exact structured predicates, makes no traditional LLM call at query time, builds no per-corpus index, and returns an interval with measured coverage.

\paragraph{Statistical estimation with model predictions.}
Prediction-powered inference (PPI)~\cite{angelopoulos2023ppi} combines a model's predictions with a few gold labels into valid intervals, even for a biased model.
For \sys{} those labels would be judge verdicts on the query's sample, which puts the judge back on the query path; we leave this to future work (Section~\ref{sec:conclusion}).
PPI under label shift~\cite{angelopoulos2023ppi} and corrected LLM-judge estimates~\cite{lee2026judge} reuse labels from another distribution, but run the model on every record and assume a known shift model.
Closest in mechanism, Xu et al.\ calibrate per-category classifiers to a production LLM labeler and report prevalence with no further LLM calls~\cite{xu2026calibrate}.
Their categories are fixed and scored in a daily batch; \sys{} serves arbitrary criteria posed at query time with one calibrator shared across tenants, over an exactly counted filtered set, in seconds.
Wilkins-Reeves et al.\ widen proxy intervals by a bias distribution learned from past domains~\cite{wilkinsreeves2026estimate}; we fit our margin by split conformal prediction across tenants and measure its coverage on held-out tenants.
In quantification terms, our point estimate is calibrated probabilistic classify-and-count~\cite{bella2010quantification,card2018calibration}.

\section{Conclusion}
\label{sec:conclusion}

\sys{} answers COUNT queries with structured and semantic predicates in seconds, without calling an oracle at query time.
It counts the filtered set exactly and draws a uniform sample in the same columnar scan.
It scores the sample with a fast classifier calibrated to the production judge's logged verdicts, and widens the interval by one-sided, rate-dependent conformal margins for calibration error.
In production, latency had a p99 of 3.76\,s on the latency workload.
On held-out tenants, the nominal 90\% interval covered the true count in 92.0\% of estimates.

Three directions remain.
A prediction-powered correction~\cite{angelopoulos2023ppi} from a few judge verdicts on the sample would remove the assumption that calibration transfers.
SUM or AVG of a structured column over the semantic matches follows as a ratio estimator on the same sample.
A bounded-time sampler over a copy of the table ordered by tenant and a fixed hash of the record id would bound the scan for the largest scopes.

\bibliographystyle{ACM-Reference-Format}
\bibliography{refs}

\end{document}